\documentclass[aps,prl,twocolumn,superscriptaddress,showpacs]{revtex4}

\usepackage{graphicx}% Include figure files
\usepackage{dcolumn}% Align table columns on decimal point
\usepackage{bm}% bold math

\def\kpar{\mathbf{k}_\parallel}
\def\Kpar{\mathbf{K}_\parallel}

\begin{document}

\title{Reversal of spin polarization in Fe/GaAs (001) driven by resonant surface states: First-principles calculations}

\author{Athanasios N. Chantis}
\thanks{achantis@lanl.gov,Theoretical Division, Los Alamos National Laboratory, Los Alamos, New Mexico 87545, USA}
\affiliation{Theoretical Division, Los Alamos National Laboratory, Los Alamos, New Mexico 87545, USA}
\author{Kirill D. Belashchenko}
\affiliation{Department of Physics and Astronomy and Nebraska Center
for Materials and Nanoscience, University of
Nebraska-Lincoln, Lincoln, Nebraska 68588, USA}
\author{Darryl L. Smith}
\affiliation{Theoretical Division, Los Alamos National Laboratory, Los Alamos, New Mexico 87545, USA}
\author{Evgeny Y. Tsymbal}
\affiliation{Department of Physics and Astronomy and Nebraska Center
for Materials and Nanoscience, University of Nebraska-Lincoln,
Lincoln, Nebraska 68588, USA}
\author{Mark van Schilfgaarde}
\affiliation{School of Materials, Arizona State University, Tempe, Arizona 85287, USA}
\author{Robert C. Albers}
\affiliation{Theoretical Division, Los Alamos National Laboratory, Los Alamos, New Mexico 87545, USA}
%draft
\date{\today}

\begin{abstract}

A minority-spin resonant state at the Fe/GaAs(001) interface is
predicted to reverse the spin polarization with voltage bias 
of electrons transmitted
across  this interface. Using a Green's function approach within the
local spin density approximation we calculate spin-dependent current
in a Fe/GaAs/Cu tunnel junction as a function of applied bias
voltage. We find a change in sign of the spin polarization of
tunneling electrons with bias voltage due to the interface
minority-spin resonance. This result explains recent experimental
data on spin injection in Fe/GaAs contacts and on tunneling
magnetoresistance in Fe/GaAs/Fe magnetic tunnel junctions.

\end{abstract}

\pacs{72.25.Mk, 73.23.-b, 73.40.Gk, 73.40.Rw}

\maketitle

%[INTRODUCTION]\\

Ferromagnetic metal/nonmagnetic semiconductor contacts have recently
attracted significant interest due to the possibility to generate
non-equilibrium electron spin distributions in normal semiconductors
and hence be practical for spintronics applications \cite{Zutic}.
The contact structures in which electron tunneling dominates
transport properties are used to achieve a sizable spin polarization
of the electric current and produce spin accumulation in the
semiconductor. The spin polarization originates from the spin dependence
of the wavefunctions
and densities of states of the ferromagnetic contact. As a result, the tunneling
transmission coefficients are different for  majority- and
minority-spin electrons.

Among various ferromagnet/semiconductor structures, Fe/GaAs contacts have been extensively studied, showing that spin-dependent
tunneling through Schottky barriers formed by delta doping is an efficient method for injecting non-equilibrium spin
distributions in a semiconductor \cite{Hanbicki02,Hanbicki03,Crowell05}. These experiments showed that in biased Fe/GaAs contacts
the net spin of electrons injected from Fe into GaAs is parallel to the magnetization of the ferromagnetic Fe electrode. This
implies that majority-spin electrons tunnel through the Schottky barrier more efficiently that minority-spin electrons.

Recently, however, Crooker \emph{et al.} \cite{Crooker} and Lou \emph{et al.} \cite{Crowell07} observed an anomalous bias
dependence of the transport spin polarization in Fe/GaAs Schottky barrier structures. They found that both the magnitude and sign
of the spin polarization depend on applied bias voltage producing either majority- or minority-spin accumulation in GaAs. Moser
\emph{et al.} \cite{Moser} observed a related phenomenon in Fe/GaAs/Fe magnetic tunnel junctions. They found that tunneling
magnetoresistance (TMR) changes sign  with bias voltage, reflecting the reversal of the transport spin polarization at the
Fe/GaAs interface. To explain the experiments by Crooker \emph{et al.} \cite{Crooker} Dery and Sham \cite{Sham} developed a
model suggesting that the reversal of spin polarization is due to localized states at the interface formed by electrostatic confinement of
doping profiles. This explanation, however, does not take into account a realistic electronic structure of the interface which is
known to be decisive for spin-dependent transport in the tunneling regime \cite{Tsymbal}. Moreover, it doesn't explain 
the experiment by Moser \emph{et al.} \cite{Moser}.

In this Letter, we demonstrate that the observed reversal of the spin polarization in Fe/GaAs(001) tunnel contacts  is intrinsic
to their interface electronic structure. The Fe/GaAs(001) interface supports a minority-spin interface band lying in the vicinity
of the Fermi energy \cite{Butler,Demchenko}. This interface band is reminiscent of the Fe(001) surface band observed
experimentally using scanning tunneling spectroscopy \cite{Strocsio}. Due to the coupling to continuum bulk states in Fe the
Fe/GaAs(001) interface band evolves into an interface resonant band and strongly contributes to the tunneling conductance. The
minority-spin character of this resonant band leads to the reversal of the spin polarization from positive to negative in the
relevant range of electron energies. This explains the experimental findings of anomalous bias dependence of the spin
polarization in experiments on spin injection \cite{Crooker, Crowell07} and TMR \cite{Moser}.

%[METHOD, COMPUTATIONAL DETAILS]\\

The results of experiments \cite{Crooker, Crowell07,Moser} reflect features of spin transmission across the Fe/GaAs(001)
interface. This is due to the transport spin polarization in tunneling geometry being largely controlled by the interface atomic
and electronic structure \cite{Tsymbal}. In the case of spin injection \cite{Crowell07,Crooker}, electrons injected from Fe into
GaAs tunnel through the GaAs barrier, then experience scattering by a defect or impurity, and further
propagate diffusively producing spin accumulation in GaAs. Since diffusive transport in a nonmagnetic material is independent
of electron spin, the spin polarization established within GaAs is entirely due to
asymmetry in the spin transmission across the Fe/GaAs(001) interface. A similar argument applies to spin extraction from GaAs
into Fe. In case of magnetic tunnel junctions, variations in TMR are expected to reveal spin polarizations of the two interfaces
\cite{stf}. However, Moser \emph{et al.} \cite{Moser} observed a reversed TMR only for those Fe/GaAs/Fe tunnel junctions in which
one interface was "ideal" epitaxial, whereas the other was either oxidized or cleaned by a $H^{+}$ plasma. Therefore, their
findings reveal features of spin transmission across the epitaxial Fe/GaAs(001) interface only.

To study spin-polarized transport across the Fe/GaAs(001) interface we consider a Fe/GaAs/Cu(001) tunnel junction with a bcc Cu
counter-electrode, which serves as a detector of spin polarization, in the spirit of Ref. \cite{chantis}. The bcc Cu electrode
has a spin-independent free-electron-like band structure and a featureless surface transmission function \cite{stf}, making it a
perfect spin detector. This implies that variations in the spin polarization of the tunneling current with bias voltage found in
the calculation performed for the Fe/GaAs/Cu(001) tunnel junction are entirely due to the changes in the spin transmission across
the Fe/GaAs(001) interface. This makes the results of our calculations relevant to experiments \cite{Crowell07,Crooker,Moser}.

The particular junction studied consists of a semi-infinite Fe electrode, 8 monolayers of GaAs barrier, and a semi-infinite bcc Cu
electrode. We consider an As-terminated interface, motivated by  the experiments on spin injection \cite{Crowell07,Crooker} where
the epitaxial Fe/GaAs interfaces were grown in As-rich environment \cite{personal}. Since intermixing of Fe and As atoms at this
interface is not energetically favorable \cite{Erwin,Demchenko}, we assume that the interface is abrupt. The small change of the
As-Ga interplane distance of about 0.14 \AA\ due to relaxation \cite{Demchenko} is not taken into account.

Calculations are performed using the Green's function representation of the tight-binding linear muffin-tin orbital (TB-LMTO)
method in the atomic sphere approximation (ASA) \cite{andersen}. We apply third-order parametrization for the Green's function
\cite{Gunnarson}. The electronic structure problem is solved within the scalar-relativistic density functional theory (DFT) where
the exchange and correlation potential is treated in the local spin-density approximation (LSDA). The conductance is calculated
using the principal-layer Green's function technique \cite{Turek,kudr00} within the Landauer-B\"uttiker approach \cite{Datta}.
Charge self-consistency is achieved before performing the transport calculations.

The spin-dependent transmission coefficient ${t^\sigma(E,\kpar)}$ is calculated for a given spin ${\sigma=\uparrow,\downarrow}$
(where $\uparrow$ and $\downarrow$ denote majority and minority spin, respectively) as a function of energy $E$ and the
transverse wave vector $\kpar$ which is conserved due to the transverse periodicity of the junction. The total transmission for a
given energy and spin is obtained by integrating over $\kpar$ within a two-dimensional Brillouin zone (2DBZ): ${T^\sigma(E)= \int
t^\sigma(E,\kpar)d^2\kpar}/(2\pi)^2$. A uniform 200$\times$200 mesh is used for the integration. The current density associated
with this transmission is obtained from $J^{\sigma}(V)=(e/h)\int_{E_{F}}^{E_{F}+eV} T^\sigma(E) dE$, where ${E_F}$ is the Fermi
energy and $V$ is the applied bias voltage. This is a reasonable approximation for small voltages considered in this work. This
definition of $J^{\sigma}(V)$ implies that for a {\it negative} voltage electrons tunnel {\it from} Fe across GaAs. The spin
polarization is defined as ${P=(J^{\uparrow}-J^{\downarrow})/(J^{\uparrow}+J^{\downarrow})}$.

%\\

\begin{figure}[tbp]
\includegraphics[width=0.45\textwidth]{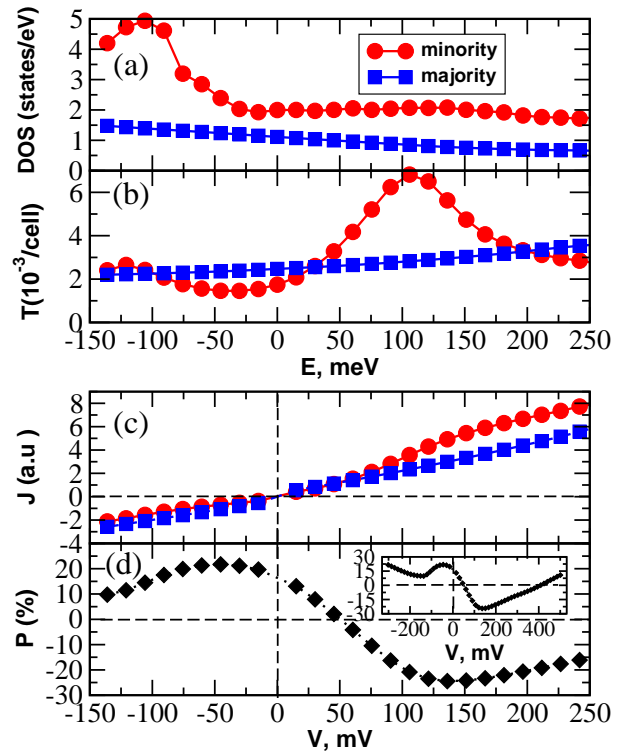}
\caption{ \small Results of calculations for a Fe/GaAs/Cu tunnel
junction: (a) spin-resolved local DOS for the interface Fe monolayer; (b)
spin-resolved integrated transmission as a function of energy; (c)
spin-resolved current density as a function of bias voltage; (d)
spin polarization as a function of bias voltage. 
The inset shows the spin polarization over an extended range of bias \cite{comment}. 
%However, this may exceed the range over which the approximations 
%used in our calculations work well, hence it is presented to show
%the qualitative picture.
  In (a) and (b), the
Fermi level is set at zero energy.  } \label{POL}
\end{figure}

%[MAIN RESULTS ACCORDING TO FIG. 1]\\

Figs. \ref{POL}a and \ref{POL}b  show the calculated local density of states (DOS) at the interface Fe monolayer and the integrated
transmission as a function of energy for the Fe/GaAs/Cu junction.
%The energy dependence of the transmission represents the
%linear-response conductance in the rigid-band model.
The energies are given with respect to ${E_F}$ which is found to be in the middle of the GaAs band gap in agreement with previous
calculations \cite{Butler,Demchenko}. As is seen from Fig. \ref{POL}a, the minority spin dominates the interface DOS in the
vicinity of  the Fermi energy throughout the energy interval shown. There is a sharp peak in the minority-spin DOS between -50
and -160 meV. The majority-spin transmission (Fig. \ref{POL}b) exhibits a featureless free-electron-like energy dependence
mirroring the featureless majority-spin DOS (Fig. \ref{POL}a). In contrast, the minority-spin transmission is nonmonotonic and
dominates in two energy windows, between $-130$ and $-110$ meV and between $+50$ and $+175$ meV (Fig. \ref{POL}b). The former
local maximum corresponds to the sharp peak in the minority-spin interface DOS, whereas the latter maximum has no distinct analog
in the DOS.

\begin{figure}[tbp]
\includegraphics[width=0.47\textwidth,clip]{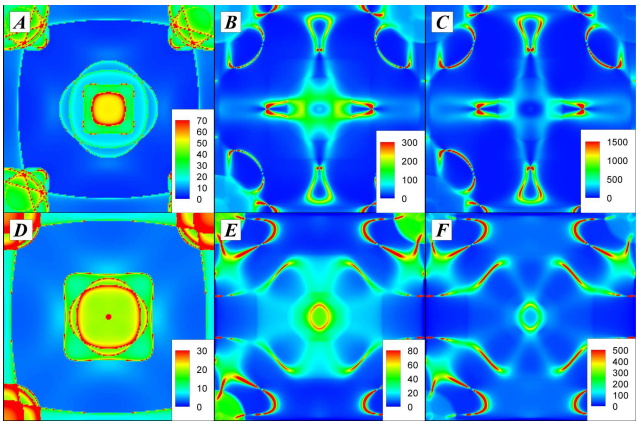}
\caption{Minority-spin Fe local density of states resolved in the
two-dimensional Brillouin zone by $\kpar$ with abscissa along
[$\bar{1}\bar{0}$] and ordinate along [$\bar{0}\bar{1}$] direction.
The upper three panels are for $E=-121$ meV corresponding to the
local maximum of the minority-spin transmission: (a) bulk, (b)
sub-interface monolayer, (c) interface monolayer. The lower three
panels are for $E=106$ meV corresponding to the maximum of the
minority-spin transmission: (d) bulk, (e) sub-interface monolayer,
(f) interface monolayer.} \label{DOS}
\end{figure}

The energy dependence of the transmission  is reflected in the voltage dependence of the spin-resolved current density shown in
Fig. \ref{POL}c. It is seen that, while for negative bias voltages majority-spin electrons dominate the current density, there is
a crossover at about $+50$ mV which makes the minority-spin current dominating at higher voltages up to $V=+400$ mV (see inset in Fig. \ref{POL}d). 
This leads to the reversal of spin polarization at about $V=+50$ mV seen in Fig. \ref{POL}d. At $V=+400$ mV the spin polarization
changes sign again reversing from anomalous (negative) to normal (positive). The transmission peak between $-130$ and $-110$ meV
(Fig. \ref{POL}b) is too small to change the sign of the spin polarization and only leads to a reduction of the spin polarization
by about 10\%. The reversal of the spin polarization with bias voltage is the central result of this Letter. In the following we
will show that an interface resonant band is responsible for this anomalous behavior.

\begin{figure}[tbp]
\includegraphics[angle=0,width=0.47\textwidth,clip]{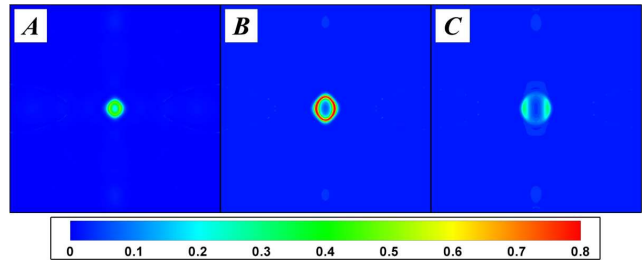}
\caption{$\Kpar$-resolved minority-spin transmission through a
Fe/GaAs/Cu (001) tunnel junction for three energies near the maximum
at $E=106$ meV: (a) 45 meV, (b) 106 meV, (c) 166 meV.} \label{TRANS}
\end{figure}

%The electronic structure of the Fe/GaAs interface is characterized
%by minority-spin interface states arising from $d_{x^2-r^2}$ and
%$d_{xy}$ orbitals on interface Fe sites that couple with the bulk Fe
%$\Delta_{2^\prime}$ minority band.

Fig. \ref{DOS} shows the $\kpar$-resolved minority-spin local DOS for two monolayers of Fe at the Fe/GaAs (001) interface in
comparison to the bulk DOS of Fe. The upper three panels correspond to the energy $E=-121$ meV at the maximum in the interface
DOS (Fig. \ref{POL}a) and the matching local peak in the transmission (Fig. \ref{POL}b). The lower three panels correspond to the
energy $E=+106$ meV at the maximum in the transmission (Fig. \ref{POL}b). It is seen that for both energies the interface DOS is
strikingly different from the respective bulk DOS (compare Figs. \ref{DOS}c and \ref{DOS}a, and Figs. \ref{DOS}f and \ref{DOS}d).
As is evident from Figs. \ref{DOS}c and \ref{DOS}f, for both energies the interface DOS are characterized by features which have
the $C_{2v}$ symmetry intrinsic to the atomic structure of the Fe/GaAs(001) interface. The topology of these features is
preserved at the sub-interface monolayer, but their intensity drops down by a factor of five (compare Figs. \ref{DOS}c and
\ref{DOS}b, and Figs. \ref{DOS}f and \ref{DOS}e). This behavior clearly points to the presence of minority-spin interface states
at energies $E=-121$ meV and $E=+106$ meV. The integral DOS for the state at $E=-121$ meV  is much higher then that for the state
at $E=+106$ meV and consequently the former produces the sharp peak in Fig. \ref{POL}a, whereas the latter is relatively broad.

The analysis of the character of the interface states (bands) shows that they arise from a mixture of $d_{3z^2-r^2}$ and $d_{xy}$
orbitals on the interface Fe sites. These states moderately hybridize with bulk Fe minority-spin bands and develop into interface
resonances. The latter fact is evident from their finite width that allows these states to be resolved in $\kpar$ space (see
Figs. \ref{DOS}c and \ref{DOS}f).

The interface resonances contribute to the tunneling conductance. The magnitude of their contribution, however, strongly depends
on their distribution across the 2DBZ, because the decay of evanescent states in GaAs depends on $\kpar$. By analyzing the complex
band structure of GaAs, Mavropoulos \emph{et al.} ~\cite{Mavr00} demonstrated that the decay constant $\kappa$ for the evanescent
states has a rather deep parabolic global minimum at the $\bar{\Gamma}$ point ($\kpar=0$). This feature strongly
suppresses the transmission through the resonant states at $E=-121$ meV, because they are located far from the $\bar\Gamma$ point
(Fig. \ref{DOS}c). In contrast, the resonance at $E=+106$ meV corresponds to the opening of a parabolic pocket at the
$\bar{\Gamma}$ point which is seen as an ellipse in the surface DOS (Fig. \ref{DOS}f); proximity to the $\bar{\Gamma}$ point
allows these electrons to tunnel efficiently across the GaAs barrier.

Fig.~\ref{TRANS} shows the $\kpar$-resolved transmission for three energies around the transmission maximum at $E=$+106 meV. It
is seen that around the $\bar{\Gamma}$ point the resonance band is parabolic and anisotropic, reflecting the $C_{2v}$ symmetry of
the interface. Owing to its location near the zone center and a relatively large DOS, this minority-spin band dominates 
the transmission near E=+106 meV. For lower energies (Fig.~\ref{TRANS}a) the resonant band only
partially crosses the Fermi level (because of its finite linewidth) providing fewer states to the tunneling current, while for
higher energies (Fig.~\ref{TRANS}c) the crossings occur for larger $\kpar$ which reduces the resonant band contribution due to
$\kpar$ filtering in GaAs. This  leads to the dominant contribution of the minority-spin resonant states in the tunneling
conductance in a finite energy window (Fig. \ref{POL}b) and results in the reversal of the spin polarization at bias voltages
from +50 to +400 mV (Fig. \ref{POL}d).

Our results explain the experimental data on spin injection by Crooker \emph{et al.} \cite{Crooker} and Lou \emph{et al.}
\cite{Crowell07}. In our calculations the reversal of the spin polarization occurs at positive applied bias voltage. This
corresponds to electrons incoming from GaAs into unoccupied states of Fe, that is {\it forward} applied bias (spin extraction) in
the experiments \cite{Crowell07,Crooker}. We find positive spin polarization for negative applied bias voltage, corresponding to
electrons incoming from Fe into GaAs, that is {\it reverse} applied bias (spin injection) in experiments
\cite{Crowell07,Crooker}. These results are in  agreement with the experimental data by Lou \emph{et al.} \cite{Crowell07}
(samples A and C). We note, however, that the energy of the interface states is sensitive to details of the sample preparation and may be
affected by inter-diffusion and disorder resulting in energy shifts of the order of several tenths of an eV. Such a shift may
explain why the reversal of the spin polarization occurs when electrons are injected from Fe into GaAs in sample B of Lou
\emph{et al.} \cite{Crowell07}.

Our results also agree with the TMR data of Moser \emph{et al.} \cite{Moser}. They observed a TMR reversal for Fe/GaAs/Fe tunnel
junctions with one epitaxial interface and the other one either oxidized or cleaned with $H^{+}$ plasma. Since no anomalies are
observed when both interfaces are disordered, the reversed TMR is entirely due to the reversal of the spin polarization at the
epitaxial interface, which occurs at bias voltages from $-90$ to about $+400$ mV. The minimum in TMR is at $V=+50$ mV which
corresponds to electrons transmitted across GaAs to the epitaxial interface.  This is consistent with our results shown in Fig.
\ref{POL}d. When a sample was annealed at $150\,^{\circ}\mathrm{C}$ for $1 \rm{h}$ the first reversal of TMR
occured at $+20$ mV instead of $-90$ mV, while the shape of the trace remained essentially unchanged \cite{Weiss}.  This supports our view
concerning the sensitivity of interface states to sample preparation. 

In conclusion, we have demonstrated that the minority-spin resonant states at the Fe/GaAs(001) interface are responsible for the
reversal of spin polarization of electrons transmitted across this interface. This explains experimental data on spin injection
in Fe/GaAs contacts and on TMR in Fe/GaAs/Fe magnetic tunnel junctions.

\begin{acknowledgments}
We thank Scott Crooker, Paul Crowell and Dieter Weiss for valuable discussions and for sharing with us their experimental
results.  The work at Los Alamos National Laboratory was supported by DOE Office of Basic Energy Sciences Work Proposal Number
08SCPE973. KDB acknowledges support from the Nebraska Research Initiative and the NSF EPSCoR First Award. 
EYT thanks the Nanoelectronics Research Initiative and NSF MRSEC for support.
MvS was supported by ONR contract N00014-07-1-0479.

\end{acknowledgments}

\bibliography{fegas}% Produces the bibliography via BibTeX.

\begin{thebibliography}{24}
\expandafter\ifx\csname natexlab\endcsname\relax\def\natexlab#1{#1}\fi
\expandafter\ifx\csname bibnamefont\endcsname\relax
  \def\bibnamefont#1{#1}\fi
\expandafter\ifx\csname bibfnamefont\endcsname\relax
  \def\bibfnamefont#1{#1}\fi
\expandafter\ifx\csname citenamefont\endcsname\relax
  \def\citenamefont#1{#1}\fi
\expandafter\ifx\csname url\endcsname\relax
  \def\url#1{\texttt{#1}}\fi
\expandafter\ifx\csname urlprefix\endcsname\relax\def\urlprefix{URL }\fi
\providecommand{\bibinfo}[2]{#2}
\providecommand{\eprint}[2][]{\url{#2}}

\bibitem[{\citenamefont{Zutic et~al.}(2004)\citenamefont{Zutic, Fabian, and
  Sarma}}]{Zutic}
\bibinfo{author}{\bibfnamefont{I.}~\bibnamefont{Zutic}},
  \bibinfo{author}{\bibfnamefont{J.}~\bibnamefont{Fabian}}, \bibnamefont{and}
  \bibinfo{author}{\bibfnamefont{S.~D.} \bibnamefont{Sarma}},
  \bibinfo{journal}{Rev. Mod. Phys.} \textbf{\bibinfo{volume}{76}},
  \bibinfo{pages}{323} (\bibinfo{year}{2004}).

\bibitem[{\citenamefont{Hanbicki et~al.}(2002)\citenamefont{Hanbicki, Jonker,
  Itskos, Kioseoglou, and Petrou}}]{Hanbicki02}
\bibinfo{author}{\bibfnamefont{A.~T.} \bibnamefont{Hanbicki}},
  \bibinfo{author}{\bibfnamefont{B.~T.} \bibnamefont{Jonker}},
  \bibinfo{author}{\bibfnamefont{G.}~\bibnamefont{Itskos}},
  \bibinfo{author}{\bibfnamefont{G.}~\bibnamefont{Kioseoglou}},
  \bibnamefont{and} \bibinfo{author}{\bibfnamefont{A.}~\bibnamefont{Petrou}},
  \bibinfo{journal}{Appl. Phys. Lett.} \textbf{\bibinfo{volume}{80}},
  \bibinfo{pages}{1240} (\bibinfo{year}{2002}).

\bibitem[{\citenamefont{Hanbicki et~al.}(2003)\citenamefont{Hanbicki, van~'t
  Erve, Magno, Kioseoglou, Li, Jonker, Itskos, Mallory, Yasar, and
  Petrou}}]{Hanbicki03}
\bibinfo{author}{\bibfnamefont{A.~T.} \bibnamefont{Hanbicki}},
  \bibinfo{author}{\bibfnamefont{O.~M.~J.} \bibnamefont{van~'t Erve}},
  \bibinfo{author}{\bibfnamefont{R.}~\bibnamefont{Magno}},
  \bibinfo{author}{\bibfnamefont{G.}~\bibnamefont{Kioseoglou}},
  \bibinfo{author}{\bibfnamefont{C.~H.} \bibnamefont{Li}},
  \bibinfo{author}{\bibfnamefont{B.~T.} \bibnamefont{Jonker}},
  \bibinfo{author}{\bibfnamefont{G.}~\bibnamefont{Itskos}},
  \bibinfo{author}{\bibfnamefont{R.}~\bibnamefont{Mallory}},
  \bibinfo{author}{\bibfnamefont{M.}~\bibnamefont{Yasar}}, \bibnamefont{and}
  \bibinfo{author}{\bibfnamefont{A.}~\bibnamefont{Petrou}},
  \bibinfo{journal}{Appl. Phys. Lett.} \textbf{\bibinfo{volume}{82}},
  \bibinfo{pages}{4092} (\bibinfo{year}{2003}).

\bibitem[{\citenamefont{Adelmann et~al.}(2005)\citenamefont{Adelmann, Lou,
  Strand, Palmstrom, and Crowell}}]{Crowell05}
\bibinfo{author}{\bibfnamefont{C.}~\bibnamefont{Adelmann}},
  \bibinfo{author}{\bibfnamefont{X.}~\bibnamefont{Lou}},
  \bibinfo{author}{\bibfnamefont{J.}~\bibnamefont{Strand}},
  \bibinfo{author}{\bibfnamefont{C.~J.} \bibnamefont{Palmstrom}},
  \bibnamefont{and} \bibinfo{author}{\bibfnamefont{P.~A.}
  \bibnamefont{Crowell}}, \bibinfo{journal}{Phys. Rev. B}
  \textbf{\bibinfo{volume}{71}}, \bibinfo{pages}{121301}
  (\bibinfo{year}{2005}).

\bibitem[{\citenamefont{Crooker et~al.}(2005)\citenamefont{Crooker, Furis, Lou,
  Adelman, Smith, Palmstrom, and Crowell}}]{Crooker}
\bibinfo{author}{\bibfnamefont{S.~A.} \bibnamefont{Crooker}},
  \bibinfo{author}{\bibfnamefont{M.}~\bibnamefont{Furis}},
  \bibinfo{author}{\bibfnamefont{X.}~\bibnamefont{Lou}},
  \bibinfo{author}{\bibfnamefont{C.}~\bibnamefont{Adelman}},
  \bibinfo{author}{\bibfnamefont{D.~L.} \bibnamefont{Smith}},
  \bibinfo{author}{\bibfnamefont{C.~J.} \bibnamefont{Palmstrom}},
  \bibnamefont{and} \bibinfo{author}{\bibfnamefont{P.~A.}
  \bibnamefont{Crowell}}, \bibinfo{journal}{Science}
  \textbf{\bibinfo{volume}{309}}, \bibinfo{pages}{5744} (\bibinfo{year}{2005}).

\bibitem[{\citenamefont{Lou et~al.}(2007)\citenamefont{Lou, Adelman, Crooker,
  Garlid, Zhang, Reddy, Flexner, Palmstrom, and Crowell}}]{Crowell07}
\bibinfo{author}{\bibfnamefont{X.}~\bibnamefont{Lou}},
  \bibinfo{author}{\bibfnamefont{C.}~\bibnamefont{Adelman}},
  \bibinfo{author}{\bibfnamefont{S.~A.} \bibnamefont{Crooker}},
  \bibinfo{author}{\bibfnamefont{E.~S.} \bibnamefont{Garlid}},
  \bibinfo{author}{\bibfnamefont{J.}~\bibnamefont{Zhang}},
  \bibinfo{author}{\bibfnamefont{K.~S.~M.} \bibnamefont{Reddy}},
  \bibinfo{author}{\bibfnamefont{S.~D.} \bibnamefont{Flexner}},
  \bibinfo{author}{\bibfnamefont{C.~J.} \bibnamefont{Palmstrom}},
  \bibnamefont{and} \bibinfo{author}{\bibfnamefont{P.~A.}
  \bibnamefont{Crowell}}, \bibinfo{journal}{Nature Physics}
  \textbf{\bibinfo{volume}{3}}, \bibinfo{pages}{197} (\bibinfo{year}{2007}).

\bibitem[{\citenamefont{Moser et~al.}(2006)\citenamefont{Moser, Zenger, Gerl,
  Schuh, Meier, Chen, Bayreuther, Wegscheider, and Weiss}}]{Moser}
\bibinfo{author}{\bibfnamefont{J.}~\bibnamefont{Moser}},
  \bibinfo{author}{\bibfnamefont{M.}~\bibnamefont{Zenger}},
  \bibinfo{author}{\bibfnamefont{C.}~\bibnamefont{Gerl}},
  \bibinfo{author}{\bibfnamefont{D.}~\bibnamefont{Schuh}},
  \bibinfo{author}{\bibfnamefont{R.}~\bibnamefont{Meier}},
  \bibinfo{author}{\bibfnamefont{P.}~\bibnamefont{Chen}},
  \bibinfo{author}{\bibfnamefont{G.}~\bibnamefont{Bayreuther}},
  \bibinfo{author}{\bibfnamefont{W.}~\bibnamefont{Wegscheider}},
  \bibnamefont{and} \bibinfo{author}{\bibfnamefont{D.}~\bibnamefont{Weiss}},
  \bibinfo{journal}{Appl. Phys. Lett.} \textbf{\bibinfo{volume}{89}},
  \bibinfo{pages}{162106} (\bibinfo{year}{2006}).

\bibitem[{\citenamefont{Dery and Sham}(2007)}]{Sham}
\bibinfo{author}{\bibfnamefont{H.}~\bibnamefont{Dery}} \bibnamefont{and}
  \bibinfo{author}{\bibfnamefont{L.~J.} \bibnamefont{Sham}},
  \bibinfo{journal}{Phys. Rev. Lett.} \textbf{\bibinfo{volume}{98}},
  \bibinfo{pages}{046602} (\bibinfo{year}{2007}).

\bibitem[{\citenamefont{Tsymbal et~al.}(2007)\citenamefont{Tsymbal,
  Belashchenko, Velev, Jaswal, van Schilfgaarde, Oleynik, and
  Stewart}}]{Tsymbal}
\bibinfo{author}{\bibfnamefont{E.~Y.} \bibnamefont{Tsymbal}},
  \bibinfo{author}{\bibfnamefont{K.~D.} \bibnamefont{Belashchenko}},
  \bibinfo{author}{\bibfnamefont{J.}~\bibnamefont{Velev}},
  \bibinfo{author}{\bibfnamefont{S.~S.} \bibnamefont{Jaswal}},
  \bibinfo{author}{\bibfnamefont{M.}~\bibnamefont{van Schilfgaarde}},
  \bibinfo{author}{\bibfnamefont{I.~I.} \bibnamefont{Oleynik}},
  \bibnamefont{and} \bibinfo{author}{\bibfnamefont{D.~A.}
  \bibnamefont{Stewart}}, \bibinfo{journal}{Prog. Mater. Science}
  \textbf{\bibinfo{volume}{52}}, \bibinfo{pages}{401} (\bibinfo{year}{2007}).

\bibitem[{\citenamefont{Butler et~al.}(1997)\citenamefont{Butler, Zhang, Wang,
  van Ek, and MacLaren}}]{Butler}
\bibinfo{author}{\bibfnamefont{W.~H.} \bibnamefont{Butler}},
  \bibinfo{author}{\bibfnamefont{X.-G.} \bibnamefont{Zhang}},
  \bibinfo{author}{\bibfnamefont{X.}~\bibnamefont{Wang}},
  \bibinfo{author}{\bibfnamefont{J.}~\bibnamefont{van Ek}}, \bibnamefont{and}
  \bibinfo{author}{\bibfnamefont{J.~M.} \bibnamefont{MacLaren}},
  \bibinfo{journal}{J. Appl. Phys} \textbf{\bibinfo{volume}{81}},
  \bibinfo{pages}{5518} (\bibinfo{year}{1997}).

\bibitem[{\citenamefont{Demchenko and Liu}(2006)}]{Demchenko}
\bibinfo{author}{\bibfnamefont{D.~O.} \bibnamefont{Demchenko}}
  \bibnamefont{and} \bibinfo{author}{\bibfnamefont{A.~Y.} \bibnamefont{Liu}},
  \bibinfo{journal}{Phys.\ Rev. B} \textbf{\bibinfo{volume}{73}},
  \bibinfo{pages}{115332} (\bibinfo{year}{2006}).

\bibitem[{\citenamefont{Stroscio et~al.}(1995)\citenamefont{Stroscio, Pierce,
  Davies, Celotta, and Weinert}}]{Strocsio}
\bibinfo{author}{\bibfnamefont{J.~A.} \bibnamefont{Stroscio}},
  \bibinfo{author}{\bibfnamefont{D.~T.} \bibnamefont{Pierce}},
  \bibinfo{author}{\bibfnamefont{A.}~\bibnamefont{Davies}},
  \bibinfo{author}{\bibfnamefont{R.~J.} \bibnamefont{Celotta}},
  \bibnamefont{and} \bibinfo{author}{\bibfnamefont{M.}~\bibnamefont{Weinert}},
  \bibinfo{journal}{Phys. Rev. Lett.} \textbf{\bibinfo{volume}{75}},
  \bibinfo{pages}{2960} (\bibinfo{year}{1995}).

\bibitem[{\citenamefont{Belashchenko et~al.}(2004)\citenamefont{Belashchenko,
  Tsymbal, van Schilfgaarde, Stewart, Oleynik, and Jaswal}}]{stf}
\bibinfo{author}{\bibfnamefont{K.~D.} \bibnamefont{Belashchenko}},
  \bibinfo{author}{\bibfnamefont{E.~Y.} \bibnamefont{Tsymbal}},
  \bibinfo{author}{\bibfnamefont{M.}~\bibnamefont{van Schilfgaarde}},
  \bibinfo{author}{\bibfnamefont{D.~A.} \bibnamefont{Stewart}},
  \bibinfo{author}{\bibfnamefont{I.~I.} \bibnamefont{Oleynik}},
  \bibnamefont{and} \bibinfo{author}{\bibfnamefont{S.~S.}
  \bibnamefont{Jaswal}}, \bibinfo{journal}{Phys. Rev. B}
  \textbf{\bibinfo{volume}{69}}, \bibinfo{pages}{174408}
  (\bibinfo{year}{2004}).

\bibitem[{\citenamefont{Chantis et~al.}(2007)\citenamefont{Chantis,
  Belashchenko, Tsymbal, and van Schilfgaarde}}]{chantis}
\bibinfo{author}{\bibfnamefont{A.~N.} \bibnamefont{Chantis}},
  \bibinfo{author}{\bibfnamefont{K.~D.} \bibnamefont{Belashchenko}},
  \bibinfo{author}{\bibfnamefont{E.~Y.} \bibnamefont{Tsymbal}},
  \bibnamefont{and} \bibinfo{author}{\bibfnamefont{M.}~\bibnamefont{van
  Schilfgaarde}}, \bibinfo{journal}{Phys. Rev. Lett.}
  \textbf{\bibinfo{volume}{98}}, \bibinfo{pages}{046601}
  (\bibinfo{year}{2007}).

\bibitem[{per()}]{personal}
\bibinfo{note}{S. A. Crooker, private communication}.

\bibitem[{\citenamefont{Erwin et~al.}(2002)\citenamefont{Erwin, Lee, and
  Scheffler}}]{Erwin}
\bibinfo{author}{\bibfnamefont{S.~C.} \bibnamefont{Erwin}},
  \bibinfo{author}{\bibfnamefont{S.-H.} \bibnamefont{Lee}}, \bibnamefont{and}
  \bibinfo{author}{\bibfnamefont{M.}~\bibnamefont{Scheffler}},
  \bibinfo{journal}{Phys. Rev. B} \textbf{\bibinfo{volume}{65}},
  \bibinfo{pages}{205422} (\bibinfo{year}{2002}).

\bibitem[{\citenamefont{Andersen}(1975)}]{andersen}
\bibinfo{author}{\bibfnamefont{O.~K.} \bibnamefont{Andersen}},
  \bibinfo{journal}{Phys. Rev. B} \textbf{\bibinfo{volume}{12}},
  \bibinfo{pages}{3060} (\bibinfo{year}{1975}).

\bibitem[{\citenamefont{Gunnarson et~al.}(1983)\citenamefont{Gunnarson, Jepsen,
  and Andersen}}]{Gunnarson}
\bibinfo{author}{\bibfnamefont{O.}~\bibnamefont{Gunnarson}},
  \bibinfo{author}{\bibfnamefont{O.}~\bibnamefont{Jepsen}}, \bibnamefont{and}
  \bibinfo{author}{\bibfnamefont{O.~K.} \bibnamefont{Andersen}},
  \bibinfo{journal}{Phys.\ Rev. B} \textbf{\bibinfo{volume}{27}},
  \bibinfo{pages}{7144} (\bibinfo{year}{1983}).

\bibitem[{\citenamefont{Turek et~al.}(1997)\citenamefont{Turek, Drchal,
  Kudrnovsk\'y, \v{S}ob, and Weinberger}}]{Turek}
\bibinfo{author}{\bibfnamefont{I.}~\bibnamefont{Turek}},
  \bibinfo{author}{\bibfnamefont{V.}~\bibnamefont{Drchal}},
  \bibinfo{author}{\bibfnamefont{J.}~\bibnamefont{Kudrnovsk\'y}},
  \bibinfo{author}{\bibfnamefont{M.}~\bibnamefont{\v{S}ob}}, \bibnamefont{and}
  \bibinfo{author}{\bibfnamefont{P.}~\bibnamefont{Weinberger}},
  \emph{\bibinfo{title}{Electronic structure of disordered alloys, surfaces and
  interfaces}} (\bibinfo{publisher}{Kluwer}, \bibinfo{year}{1997}).

\bibitem[{\citenamefont{Kudrnovsk\'y et~al.}(2000)\citenamefont{Kudrnovsk\'y,
  Drchal, Blaas, Weinberger, Turek, and Bruno}}]{kudr00}
\bibinfo{author}{\bibfnamefont{J.}~\bibnamefont{Kudrnovsk\'y}},
  \bibinfo{author}{\bibfnamefont{V.}~\bibnamefont{Drchal}},
  \bibinfo{author}{\bibfnamefont{C.}~\bibnamefont{Blaas}},
  \bibinfo{author}{\bibfnamefont{P.}~\bibnamefont{Weinberger}},
  \bibinfo{author}{\bibfnamefont{I.}~\bibnamefont{Turek}}, \bibnamefont{and}
  \bibinfo{author}{\bibfnamefont{P.}~\bibnamefont{Bruno}},
  \bibinfo{journal}{Phys. Rev. B} \textbf{\bibinfo{volume}{62}},
  \bibinfo{pages}{15084} (\bibinfo{year}{2000}).

\bibitem[{\citenamefont{Datta}(1995)}]{Datta}
\bibinfo{author}{\bibfnamefont{S.}~\bibnamefont{Datta}},
  \emph{\bibinfo{title}{Electronic transport in mesoscopic systems}}
  (\bibinfo{publisher}{Cambridge University Press}, \bibinfo{year}{1995}),
  \bibinfo{note}{ch. 3}.

\bibitem[{com()}]{comment}
\bibinfo{note}{We note that the bias interval shown here may exceed the range
  over which the approximation of small bias works well and hence the inset
  presents a trend rather than a quantitative description.}

\bibitem[{\citenamefont{Mavropoulos et~al.}(2000)\citenamefont{Mavropoulos,
  Papanikolaou, and Dederichs}}]{Mavr00}
\bibinfo{author}{\bibfnamefont{P.}~\bibnamefont{Mavropoulos}},
  \bibinfo{author}{\bibfnamefont{N.}~\bibnamefont{Papanikolaou}},
  \bibnamefont{and} \bibinfo{author}{\bibfnamefont{P.~H.}
  \bibnamefont{Dederichs}}, \bibinfo{journal}{Phys. Rev. Lett.}
  \textbf{\bibinfo{volume}{85}}, \bibinfo{pages}{1088} (\bibinfo{year}{2000}).

\bibitem[{Wei()}]{Weiss}
\bibinfo{note}{Dieter Weiss, private communication}.

\end{thebibliography}

\end{document}